# Improving the RPC rate capability

**G.Aielli**[a,b]**, P.Camarri**[a,b]**, R.Cardarelli**[b]**, A. Di Ciaccio**[a,b]**, L. Di Stante**[a,b]**, R. Iuppa**[e]**, B. Liberti**[b]**, L. Paolozzi**[d]**, E.Pastori**[b]**, R.Santonico**[a,b*]**, M.Toppi**[c]

[a] *University of Rome "Tor Vegata",*
  *Viale della Ricerca Scientifica 1, 00133 Rome Italy*

[b] *INFN "Tor Vergata",*
  *Viale della Ricerca Scientifica 1, 00133 Rome Italy*

[c] *INFN LNF,*
  *Via Enrico Fermi 40, 00044 Frascati Italy*

[d] *University of Geneva,*
  *24 rue du Général-Dufour CH 12111 Geneva 4*

[e] *University of Trento TIFPA,*
  *Via Sommarive 14, 38123 Povo, Trento, Italy*

*E-mail*: Rinaldo.Santonico@roma2.infn.it

ABSTRACT: This paper has the purpose to study the rate capability of the Resistive Plate Chamber, RPC, starting from the basic physics of this detector. The effect of different working parameters determining the rate capability is analysed in detail, in order to optimize a new family of RPCs for applications to heavy irradiation environments and in particular to the LHC phase 2. A special emphasis is given to the improvement achievable by minimizing the avalanche charge delivered in the gas. The paper shows experimental results of Cosmic Ray tests, performed to study the avalanche features for different gas gap sizes, with particular attention to the overall delivered charge. For this purpose, the paper studies, in parallel to the prompt electronic signal, also the ionic signal which gives the main contribution to the delivered charge. Whenever possible the test results are interpreted on the base of the RPC detector physics and are intended to extend and reinforce our physical understanding of this detector.

---

[*] Corresponding author.

# Contents

## 1. Introduction

The RPC is a unique gaseous detector due to the absence of any drift time. This means that the full gas volume is critical i.e. the field is strong enough to produce avalanche multiplication as soon as a free electron is generated in any point of the gas. It is therefore an ideal detector for fast decisions and very high time resolutions. Moreover, its simple structure and low cost materials make it suitable for very large area applications. This makes the RPC a very suitable detector for triggering and for time of flight measurements at present and future colliders, in particular the High Luminosity LHC. In the latter case however the required rate capability might be beyond the limits of the RPC generation presently working at LHC [ref 1] and a substantial improvement is required to overcome this limit. At the same time, it has to be stressed that any rate capability improvement requires also a careful analysis of the detector ageing, to avoid that the improved rate causes a shortened detector lifetime. This paper studies adequate methods to improve the RPC rate capability without increasing the ageing.

The RPC rate capability is mainly limited by the current that can flow through the high resistivity electrodes and can be improved working on a number of interconnected parameters that are discussed below [ref 2].

The current density flowing across the electrodes can change from point to point, on a few millimetre scale, depending on the map of the points where the gas was ionized. In average however, for a uniform irradiation density, a working current $i$ flowing across the electrodes of total resistance $R$ produces a voltage drop

$$V_{el} = V_a - V_{gas} = Ri$$

$V_a$ and $V_{gas}$ being the voltages applied to the detector and to the gas gap respectively. A high rate capability requires to keep the difference $V_{el}$ at a negligible value wrt $V_{gas}$ even under heavy irradiation. The voltage drop across the electrodes can be expresses as a function of the basic detector parameters with the following relationship

$$V_{el} = \rho\, t\, <Q>\, r_u$$

Here $\rho$ is the bulk resistivity of the electrode material; $t$ is the total thickness of both electrodes; $<Q>$ is the average charge delivered in the gas for each count and $r_u$ is the counting rate per unit surface. We define here the "rate capability" as the achieved rate per unit voltage drop across the electrodes:

$$Rate\ Capability = r_u/V_{el} = \frac{1}{\rho\, t\, <Q>}$$

It can be increased by decreasing each of the 3 parameters $\rho$, $t$, $<Q>$, with the advantages/disadvantages discussed in the following.

The search of low resistivity materials for the electrode plates is the most natural way of increasing the rate capability. However, a rate improvement obtained only by lowering the electrode resistivity, would imply a corresponding increase of the operating current and of the pollutants generated inside the gas. Moreover, the spontaneous detector noise could increase as well. This focusses a potential ageing problem and requires a new robust ageing test to qualify the RPCs for a high current working mode.

To reduce the electrode thickness $t$ has in principle a similar effect on the rate capability as to reduce its resistivity, the total electrode resistance being proportional to the product $\rho\, t$. However, a further important effect of the electrode thickness, is to attenuate the charge induced on the read out electrodes by a factor of

$$A = 1 + \frac{t}{\varepsilon_r g}$$

[ref 3]. Here $\varepsilon_r$ and $g$ are the relative dielectric constant of the electrode material and the gas gap size respectively. A thinner electrode "amplifies" therefore the signal induced on the read out electrodes, thus improving the signal to noise ratio. This effect, which is particularly relevant for very thin gaps, shows that the electrode thickness should be reduced according to the gap size.

The approach of reducing the average charge $<Q>$ has the advantage to improve the achievable rate with minimum or null penalization of other parameters such as the detector ageing. The most ambitious goal, in this perspective, is to increase the rate at constant current, thus making any further ageing test unnecessary. This approach requires: i) to optimize the gas gap size in order to find the best compromise between detection efficiency and delivered charge; ii) to transfer, from the gas to the front end electronics, a relevant part of the amplification needed to get a detectable avalanche signal. The circuit required for this purpose must be very sensitive, with an excellent signal to noise ratio. Moreover, it requires also a careful optimization of the chamber structure as a Faraday cage, to achieve a high level suppression of the noise originated both by the detector itself and by external sources. The limit of this approach is just the achievable signal to noise ratio.

This paper has the purpose to investigate how the RPC rate capability can be improved, following a pure "low charge" approach, based on a systematic study of the performance vs the gap size, under the constraints imposed by the front end electronics gain and signal to noise ratio. This approach is particularly suitable for the RPCs built with electrodes of phenolic High Pressure Laminate (HPL), whose resistivity, of about $\sim 3*10^{10}\ Ohm \times cm$, [ref 1] is already sufficiently low, with respect to other materials used so far. Electrodes of this material have been used for the largest and most relevant RPC detectors in operation now or in the recent past: Atlas, CMS, Alice Trigger, Opera, Argo-YBJ.

## 2. Study of the signals generated by RPCs of different gap sizes

We report here a Cosmic Ray laboratory test performed on three RPCs with different gap sizes, 0.5 mm, 1 mm and 2 mm respectively, but identical for any other aspect. The gas volumes, of active area 52.6 x 7.6 cm$^2$, are built with the same materials as those operating in Atlas [ref 4] and in particular with the same electrodes of phenolic High Pressure Laminate, 1.8 mm thick. The working gas is a mixture of Tetra-Fluor-Ethane (commercial name Suva 134 a), iso-Butane and Sulphur Hexafluoride: $C_2H_2F_4$/i-$C_4H_{10}$/$SF_6$ = 94.5/5.0/0.5, and is also almost identical with the standard Atlas and CMS gas. The chamber lay out, specifically conceived for this test, is shown in figure 1.

The Cosmic Rays crossing the detector under test are selected by the coincidence of other three RPCs, focussing a chamber area of 50 x 3.0 cm$^2$ at the centre of the tested RPC. This area is covered by a single read out strip shaped as shown in figure 2, which behaves like a 50 Ohm transmission line terminated at both ends with its impedance, compatible with the standard 50 Ohm coaxial cables.

The prompt signal charge mentioned in this work is defined as the charge induced on the read out copper strip by the avalanche electrons drifting in the gas. It is corrected taking into account that the signal read out by the scope at one end of strip brings only half of the primary charge, the other half being absorbed by the resistor matching the line at the opposite end.

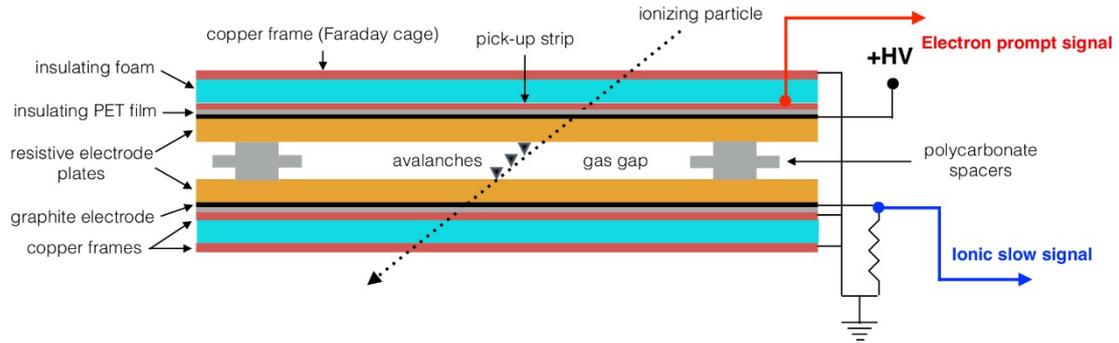

**Figure 1 Sketch of the RPCs under test.**

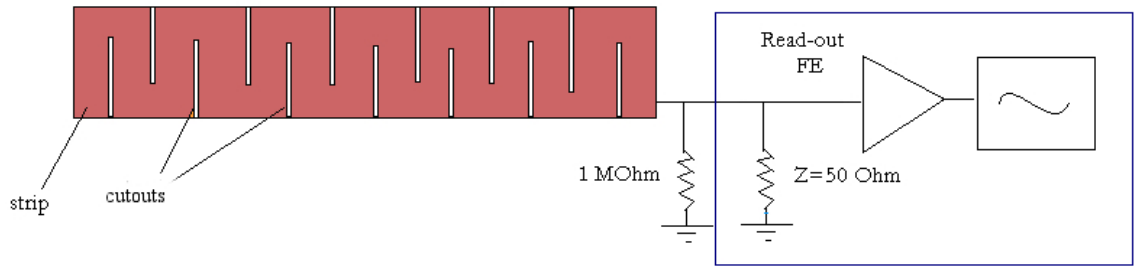

**Figure 2 The shape of the read out strip. This shape allows to cover a 2.7 cm width with relatively high impedance line, about 50 Ohm, well matched with the standard coaxial cables used to send the signals to the oscilloscope.**

The experimental set up includes also: the high and low voltage systems, the gas mixing station, a few logic units to elaborate the trigger signal and the Data Acquisition. For each trigger signal, the following waveforms are recorded on two different oscilloscopes:
   a) the prompt signal, induced on the read out strip by the electrons drifting inside the gas gap. This signal is recorded with a sampling rate of 10 Gsample/s and 1 GHz analog band
   b) the slow signal, induced on the graphite electrode by the ions drift. It is read out on a 10 kOhm resistor introduced in the HV circuit as shown in fig 1 and recorded with 5 Msample/s and 1 GHz analog band.

In both cases the signals from the RPC under test are directly fed into the oscilloscope and the recorded waveforms for the three gaps are shown in figure 3.

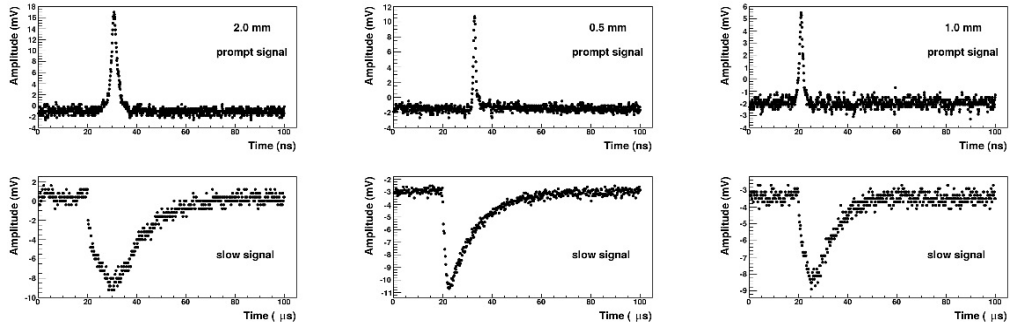

**Figure 3 Wave forms generated by (from left to right): 2 mm gap, 1 mm gap, 0.5 mm gap respectively. The prompt signals due to the electron drift on top and the slow signals due to the ion drift on the bottom are shown for each trigger. Note the horizontal scale, in nano and micro-second respectively.**

The comparison of the fast and slow signal rise times, for all three gaps, shows a constant ratio around $5 \times 10^3$, consistent with the drift velocities of electrons and ions respectively.
The comparison of waveforms from different gaps shows that the rise times of both prompt and ionic signals scale proportionally to the gap size, suggesting that the drift velocities of electrons and ions are both field independent.
The prompt signals are recorded under two different conditions: a) the signal is directly feed into the oscilloscope, as shown above; b) the signal is previously amplified with a fast charge amplifier [ref 5] and subsequently sent to the oscilloscope.  These two kinds of signal are referred, in the following, as amplified and non-amplified respectively. The ionic signal, recorded with both amplified and not amplified prompt pulses, is used to correlate them. For each working condition, the recorded waveforms are analyzed to measure all relevant parameters, such as signal charge and amplitude, signal duration and timing.
The features of the charge amplifier used in the test, are reported in table 1

| Sensitivity | 2 mV/fC |
|---|---|
| Noise | 0.3 fC RMS |
| Latch capability | 100 ps |
| Bandwidth | 50 MHz |
| Power consumption | 6 mW |
| $V_{th}$ | > 15 mV |
| $Q_{th}$ | > 6 fC |

Table 1.  Main features of the charge amplifier used in the test

The amplifier calibration plot, shown in figure 4, gives the output amplitude measured for different input charges. The points shown in this plot are obtained by binning the ionic charge in bins of 0.4 pC and correlating, in each bin, the average output amplitude of the amplified signal with the average input charge of the not amplified signal. The plot shows that the amplifier output saturates at 800 mV and, for small signals, the ratio of the output-amplitude/input-charge is about 200 mV/0.1 pC =2 mV/fC. The supplied front end voltage is 5 V.

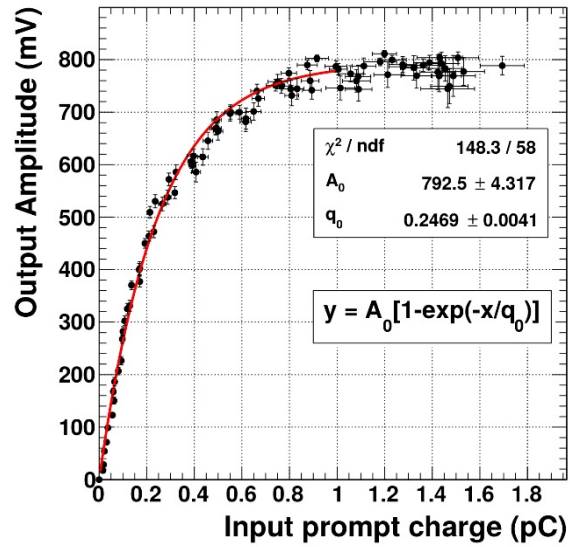

**Figure 4 Front end amplifier calibration plot.**

The detection efficiency can be measured using both the prompt and the ionic signal, with appropriate threshold criteria to discriminate the signal from the noise. These criteria are listed below:
  a) The amplitude of the non-amplified prompt signal exceeds a fixed threshold of 1.5 mV (corresponding to a charge about 150 fC for a 2 mm gap). This threshold is equivalent to that of the RPCs working in Atlas.
  b) The amplitude of the amplified prompt signal exceeds a threshold of 40 mV. According to the calibration plot of figure 4 this corresponds to a 20 fC charge threshold for the input signal.
  c) For the ionic signal instead, the threshold in amplitude is normalized to the output noise: $V_{thresh} = 5\sigma_{noise}$, where $\sigma_{noise}$ is the noise standard deviation.

The detection efficiencies achievable with the prompt signal, for the three different gaps, are shown in figure 5 vs the applied field. This figure allows to compare the efficiencies achievable with both amplified and non-amplified signals.

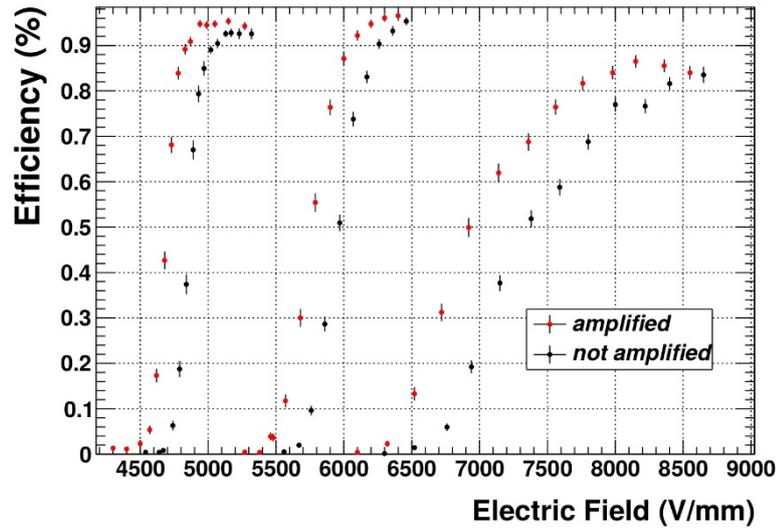

**Figure 5 Plots of the detection efficiencies vs the applied electric field, for the amplified (red points) and not amplified (black points) signals. The plots refer to 2 mm, 1 mm and 0.5 mm gas gaps respectively.**

The comparison shows that:
1) The operating field strongly increases for decreasing gap size.
2) The amplified efficiency plots, compared to the not amplified ones, are systematically shifted towards a lower field. For the 2 mm gap, e g, the voltage shift is ΔV = -350 V. The avalanche charge is consequently lowered by a factor of about 3 (see figure 6) at fixed efficiency and the rate capability should be improved accordingly.
3) The plateau efficiency of the 0.5 mm gap is substantially lower, about -10%, with respect to the gaps of 2 mm and 1 mm, indicating an insufficient primary ionization

Figure 6 shows the plot of the prompt signal amplitude vs the applied field for the three gaps. The corresponding prompt charge is plotted in figure 7. The comparison of the plots shows that just above of the plateau knee the amplitudes are approximately the same, about 3 mV, for all three gaps, as expected by the selected threshold condition, whereas the charges scale proportionally with the gap size.

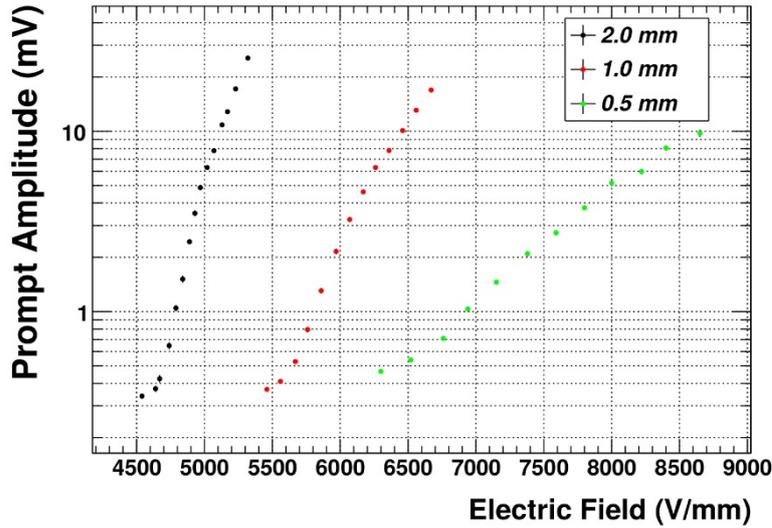

**Figure 6** Prompt signal amplitude vs electric field for the three gaps.

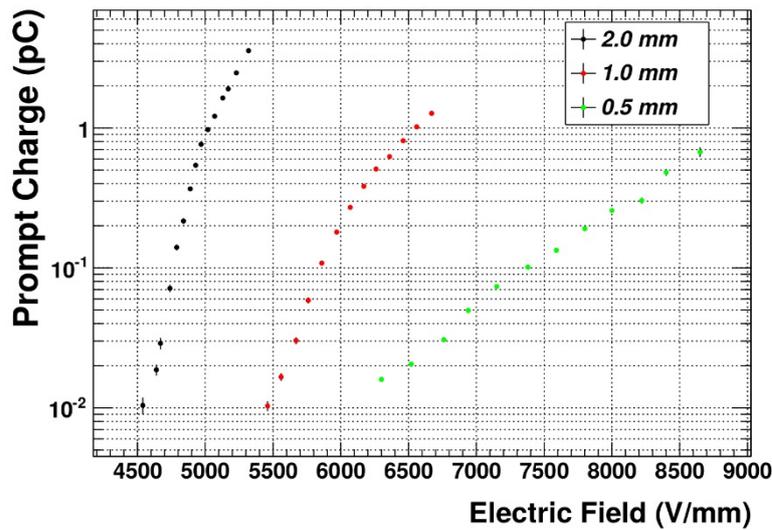

**Figure 7** Prompt signal charge vs the electric field for the three gaps.

The overall result of these tests is that thinner gas gaps reduce the delivered charge linearly with the gap size and improve the rate capability accordingly. The detection efficiency however tends to decrease due to the smaller number of primary ionizations produced in the gap. This number can be evaluated assuming a model in which the efficiency is given by the ionizations occurring in the "effective gap", which is a convenient fraction of the gas gap on the cathode side. Its width increases with the applied voltage. The inefficiency, $1 - \varepsilon$, is the "zero probability" of a Poisson distribution with <n> ionizations in average [ref 6].

$1 - \varepsilon = e^{-<n>}$ → $<n> = ln(\frac{1}{1-\varepsilon})$.

The value of <n>, which is proportional to the effective gap size, is plotted in figure 8 vs the applied voltage, for the 1 mm gas gap. The linear fit is also reported. At 6560 V in the plateau, e.g. the intrinsic inefficiency is 1.5% with 4.2 primary ionizations in average.

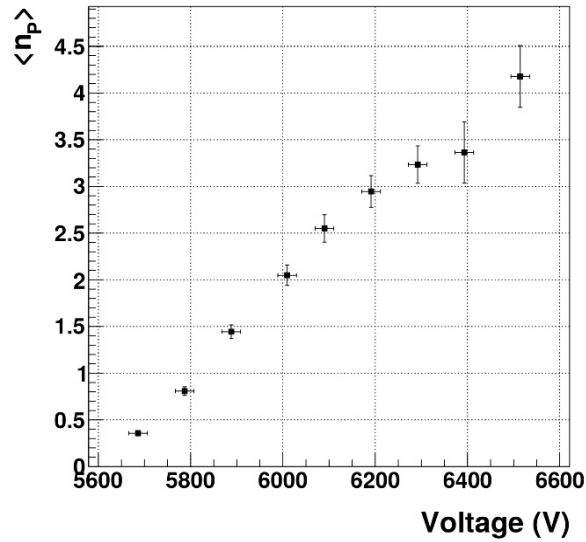

**Figure 8 Number of primary ionizations contributing to the detection efficiency vs the applied voltage. The plot refers to the 1 mm gap.**

The ionic charge, shown in fig 9, is normally much larger than the prompt charge [ref 3].

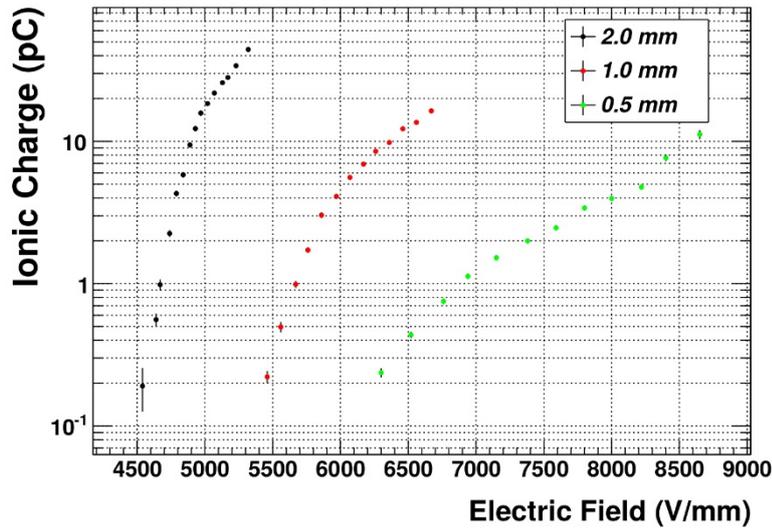

**Figure 9 Ionic charge vs electric field.**

A further comparison parameter for different gaps is the ratio Q/q of the ionic (Q) to the prompt (q) charge. Fig 10 shows this ratio vs the ionic charge Q for the three gap sizes. The three plots have very similar shapes, with a maximum for small Q, in the range of very few pC, and an asymptotic value for large Q values. The maxima slightly grow for increasing gap, suggesting that, at low voltages, the prompt charge collection is more efficient for thinner gaps. The asymptotic values are the same, about 12, for all three gaps.

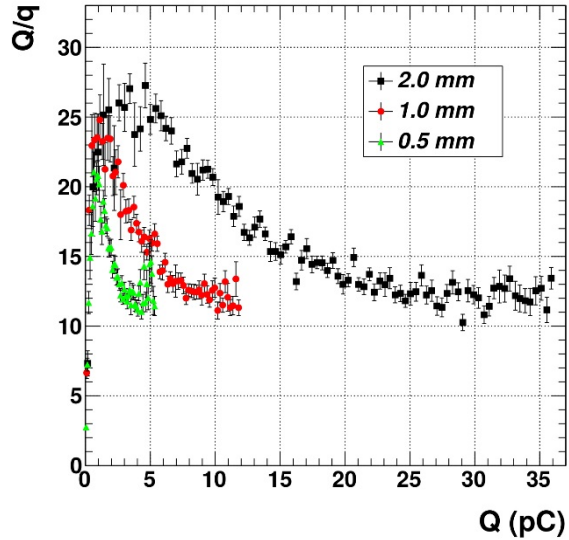

**Figure 10 Ratio of the ionic to prompt charge, Q/q vs the ionic charge Q.**

## 3. Combined effect of thinner gaps and higher sensitivity front end electronics

So far we studied, in an independent way, the effects of thinner gas gaps and of the new front end electronics with improved sensitivity. Both effects contribute to reduce the delivered charge and therefore to improve the rate capability. However, when the two effects are combined, the overall gain depends on their correlation and could range, in principle, between two extreme cases:
i) the two effects are uncorrelated (the most favourable case) and the overall gain is the product of the two independent gains;
ii) ii) the gain due to front end electronics embodies that of the thinner gap, which do not give any further advantage (the most unfavourable case).

This section shows that the real case is in between. A comprehensive presentation of the delivered charge, under all tested conditions, is summarized in table 2, which shows the results of the three gap sizes with and without the front end amplifier.

Table 2

| Gap size (in mm) | Efficiency @ plateau knee | Op.Voltage Not Amplified Signal | Average Charge (Prompt+Ionic) Not Amplified Signal | Op.Voltage Amplified Signal | Average Charge (Prompt+Ionic) Amplified Signal |
|---|---|---|---|---|---|
| 2 | 90% | 10040 V | 19.6 pC | 9660 V | 6.5 pC |
| 1 | 90% | 6260 V | 9.3 pC | 6010 V | 4.9 pC |
| 0.5 | 77% | 4000 V | 4.3 pC | 3780 V | 3.05 pC |

Table 2. Average value of the prompt+ionic charge delivered by 2 mm, 1 mm and 0.5 mm gaps at the efficiency plateau knee. The columns 3 and 4 refer to the output signal directly feed into the oscilloscope. In the columns 5 and 6 the signal was previously amplified with the circuit described in table 1.

Table 2 allows to factorize the effects of the gas gap and the front end amplification. It shows that:
1) When the gap is reduced from 2 mm to 0.5 mm in absence of amplification the delivered charge decreases by a factor 19.6/4.3 = 4.6
2) When the signal is amplified at constant gap size of 2 mm the delivered charge decreases by a factor of 19.6/6.5 = 3.0.
3) The combined effect in absence of correlation would give a factor 4.6x3 = 13.8
4) In the real case, due to the correlation, the overall gain in reducing the delivered charge is a factor of 19.6/3.05 = 6.4

The advantage of thin gaps, in addition to an improved front end sensitivity, is due to the fact that the higher achievable field produces a faster avalanche saturation thus reducing the large amplitude spread of the signals generated inside the gas.

## 4. Conclusions

We performed a systematic study of the parameters defining the charge delivered in the RPC gas for each output signal. Reducing this charge indeed is crucial to improve the RPC rate capability. We have shown that the output signal duration is proportional to the gap and therefore, for thinner gaps, the delivered charge is reduced accordingly. The combined effect of reducing the gap size from 2 mm to 0.5 mm and improving the front end electronics thanks to a new charge amplifier with 2 mV/fC sensitivity and 0.3 fC rms noise, is an overall gain of a factor 6.4 in the delivered charge and therefore in rate capability. We have also demonstrated that the operating field strongly increases for decreasing gap size and that the ratio of the ionic to prompt charge is asymptotically the same for all gaps with a numerical value of about 12.